\documentclass[11pt,a4paper,fleqn]{cas-sc}
\usepackage[authoryear,round]{natbib}

\usepackage{gensymb}
\usepackage[section]{placeins}
\usepackage{algpseudocode}
\usepackage{float}
\usepackage{wasysym}
\usepackage{longtable}

\usepackage{multirow}

\usepackage[linesnumbered,lined,boxed,commentsnumbered,ruled,longend]{algorithm2e}
\usepackage{amsmath,mathtools}

\newcommand{\RN}[1]{%
  \textup{\uppercase\expandafter{\romannumeral#1}}%
}

\begin{document}

\let\WriteBookmarks\relax
\def\floatpagepagefraction{1}
\def\textpagefraction{.001}
\shorttitle{Surrogate-based prioritization of sub-problems}
\shortauthors{Yu, Jürgens, and Göke}

\title [mode = title]{Surrogate-based prioritization of sub-problems for Benders decomposition in energy planning}                  

\author[1]{Wanhong Yu}[type=editor,auid=000,bioid=1]
\author[2]{Boyung Jürgens}[type=editor,auid=000,bioid=1]
\author[1,3]{Leonard Göke}[type=editor,auid=000,bioid=1]

\cormark[1]

\affiliation[1]{
             organization={Reliability and Risk Engineering, Institute of Energy and Process Engineering, ETH Zurich},
             addressline={8092}, 
             city={Zurich},
             country={Switzerland}}

\affiliation[2]{
             organization={Institute of Technical Thermodynamics, RWTH Aachen University},
             addressline={52062}, 
             city={Aachen},
             country={Germany}}

\affiliation[3]{
             organization={Energy and Process Systems Engineering, ETH Zurich},
             addressline={8092}, 
             city={Zurich},
             country={Switzerland}}

\cortext[cor1]{Corresponding author.}

\begin{abstract}
Benders decomposition solves optimization problems by separating the first-stage master problem from one or more second-stage sub-problems. While the standard Benders decomposition solves all sub-problems in each iteration, solving only selected sub-problems still guarantees convergence and can reduce solution time, but raises the question of how to select.

In this work, we introduce surrogate-based prioritization of sub-problems. The method leverages surrogates to estimate the sub-problems' objectives, assess the current error of the cutting-plane estimator, and then prioritize the sub-problem with the largest error. We implement surrogate-based prioritization within sequential and asynchronous Benders decomposition. Both these algorithms also leverage the surrogate to trigger convergence checks and implement regularization.

Benchmarks for an energy planning problem with a few large sub-problems show that the applied prioritization strategy works. The reduction in solution time correlates with the surrogate's accuracy. In our case, geometric interpolation-based surrogates are more accurate than machine learning methods. As a result, prioritization consistently and significantly outperforms the standard algorithm in sequential Benders decomposition. The speed-up increases with the number of scenarios, reaching 33\% with four scenarios and 55\% with ten scenarios. In the case of asynchronous parallelization, the impact on performance is less clear, and the average speed-up from prioritization is 19\%.
\end{abstract}


\maketitle

\section{Introduction}

Capacity expansion models are key tools for energy planning, but the transition from fossil fuels to renewable sources challenges established modeling practices \citep{Goeke2021}. Unlike fossil fuels, renewable sources fluctuate across daily, seasonal, and inter-annual scales, requiring models to incorporate higher temporal resolutions and diverse weather scenarios to ensure system reliability \citep{StefanPfenninger.2017}. In addition, renewable systems depend on short- and long-term storage to balance fluctuations, imposing dependencies across the modeled time-steps \citep{Sepulveda2021}. Overall, linear optimization problems in energy planning are growing in size and complexity, making them intractable for off-the-shelf solvers.

\subsection{Benders Decomposition in energy planning}

Benders decomposition (BD) solves linear optimization problems by decomposing the initial problem into a master problem (MP) for capacity expansion in the first stage, and several sub-problems (SPs) for operation in the second stage \citep{Benders1962}. Then, the algorithm iteratively determines the optimal value of the complicating variables that connect the MP and SPs by constructing a cutting-plane approximation of the SPs within the MP. Initial applications of BD in energy planning date back to 1988 \citep{firstenergybenders}. However, these applications do not leverage BD to address the challenges arising from renewables and storage in energy planning. Most previous works omit energy storage and the temporal dependencies it imposes between the different SPs for operation \citep{Lohmann.2017}. Instead, BD addresses problems that are combinatorially complex due to integer variables \citep{CristianaL.Lara.2018,CanLi.2022}. 

Against this background, several recent studies introduced BD for planning future systems with renewables and storage. Here, BD does not address combinatorial complexity but the great size of the underlying optimization problem and the dependencies imposed by storage variables. A refinement critical to the performance of these Benders applications is regularization to mitigate the heavy oscillations of BD \citep{Goke.2024,Pecci.2025,Wu2025}. In addition, applications benefit from reformulations that introduce complicating variables for storage to decompose the SPs \citep{Jacobson2024, Sasanpour2025} or from more advanced decomposition techniques \citep{Parolin2026, Gruebler2025}. Finally, generating inexact cuts by not solving SPs to optimality and parallelizing SPs accelerates convergence \citep{zakeri2000inexact,Goke.2024}. However, parallelization requires large distributed memory resources and can be inefficient \citep{Goke.2024}.

Other refinements for BD proposed in the literature are not applicable, as they address problems with a difficult mixed-integer MP and small SPs \citep{Rahmaniani.2017}. In energy planning, the SPs for operation are typically large due to temporal detail and storage dependencies, whereas the MP is typically small and continuous. The high temporal resolution and interdependencies of storages result in large SPs for system operation, while the MP for capacity expansion is small and often continuous.

\subsection{Contribution}

In this paper, we introduce surrogate-based prioritization as a refinement to speed up BD by prioritizing certain SPs instead of solving all SPs in each iteration. For this purpose, surrogates of the SPs estimate their objective to determine its priority. In contrast, previous work prioritized SPs solely based on input data using sample-average approximations, which is limited to cases where SPs are generated by sampling uncertain parameters from a known distribution and does not apply to problems with a few discrete scenarios \citep{Kothari.2026, Bertsimas.2025}.

We leverage the surrogate-based prioritization to develop a sequential and two asynchronous parallelized variants of BD. To implement regularization and check convergence, these algorithms also rely on the surrogate for estimating the SPs objective.

Our work contributes to the emerging research on combining decomposition methods with surrogates. \citet{Mazzi.2021} introduce surrogates that yield inexact cuts for unsolved SPs based on a few solved SPs; \citet{Mana.2023} replace a complex MP with a surrogate to propose new candidate solutions; and \citet{Borozan.2024} rely on a surrogate to select the most effective cuts for BD. 

The structure of the remainder of the paper is as follows: Section \ref{base} introduces the two-stage energy planning problem, the standard BD, and regularization. Then, Section \ref{ref2} introduces the surrogate-based prioritization and its implementation into sequential and asynchronous BD algorithms. Section \ref{case} introduces a case study for benchmarking the developed method in the following Section \ref{bench}. The final Section concludes.

\section{Benders algorithm for energy planning} \label{base}

The two-stage energy planning problem determines capacities in the first stage and operations in the second. Scenarios in the second stage incorporate uncertainties, for instance, weather conditions and fuel prices. Expansion decisions are taken in the first stage before the realization of the uncertainty. Second-stage operation depends on the expanded capacity determined in the first stage. The following subsections first introduce the closed-formulation of the problem, followed by the established regularized BD methods applied to solve it. 

\subsection{Problem formulation} \label{form}

We introduce the standard compact formulation (Eqs. \ref{eq:1a} to \ref{eq:1d}) to describe the two-stage stochastic problem where $x$ represents the expansion variables and $y$ represents the operation variables. In energy planning, expansion variables encompass the expansion of technology capacity and transmission capacity between regions. Operation variables describe the operation of different technologies and the energy exchange among regions at each time step. Eq. \ref{eq:1b} and Eq. \ref{eq:1d} outline constraints for expansion and operation problems, respectively. In energy planning, an example of a constraint in the expansion problem is an upper capacity limit; constraints in the operation include balancing supply and demand at every time step, tracking storage levels over time, and restricting operations to the expanded capacities (Eq. \ref{eq:1d}). 

{
\begin{subequations}
\begin{alignat}{4}
\min_{x,y_s \geq 0} c^T x &+ \sum_{s \in S} z_s && \label{eq:1a}\\
s.t. \ Hx &\leq a && \label{eq:1b}\\
z_s &= \pi_s d^T y_s && \label{eq:1c}\\
Ix+Jy_s &\leq b && \label{eq:1d}
\end{alignat}
\end{subequations}
}

The problem minimizes the sum of the first-stage expansion cost and the expected operational costs across all scenarios in the second stage (Eq. \ref{eq:1a}). The parameter $c$ denotes the costs of first-stage decisions, and $d$ denotes the costs of second-stage decisions. The first decisions are subject to constraints expressed by the matrix $H$ and the vector $a$. The variable $z_s$ denotes the weighted second-stage objective of each scenario $s$ according to their known probability $\pi_s$ (Eq. \ref{eq:1c}). The matrices $I$ and $J$, and the vector $b$ reflect the scenario-dependent uncertainty in the second stage. \citep{Goke.2021} provides a comprehensive formulation of the planning problem.

\subsection{Standard Benders decomposition} \label{bdStd}

BD is well-suited for two-stage stochastic problems, since the variables split into first-stage expansion variables, the complicating variables, and second-stage operation variables. The second-stage operation can be decomposed further into several independent SPs, each corresponding to a scenario.

For standard BD, the introduced planning model described by Eqs. \ref{eq:1a} to \ref{eq:1d} is  decomposed into one MP (Eq. \ref{eq:MP}) and a group of independent SPs (Eq. \ref{eq:SP}). The MP is a relaxed form of the original problem reformulated in terms of the expansion variables $x$. Constraints on second-stage variables are removed and, as a substitute, a cutting-plane estimator $\tilde{z}_s(x)$ is introduced for each second-stage scenario, indicating the approximated lower bound of the second-stage objective $z_s$ in terms of $x$. 

\begin{align} \label{eq:MP}
    (\mathrm{MP}^0)\quad \min_{x \geq 0} \ c^T x  + \sum_{s \in S}\tilde{z}_s(x) \quad
s.t. \ Hx \leq a 
\end{align}

In each iteration, the MP determines a candidate solution for the expansion variables $x$ and passes their value to the SPs. An SP is the inner minimization problem for a fixed $x$. Each SP corresponds to a second-stage scenario and is solved independently to derive the second-stage objective $z_s$ for that scenario.

\begin{align} \label{eq:SP}
    (\mathrm{SP}_s)\quad \min_{y_s \geq 0} z_s \quad
s.t. \ z_s = \pi_s d^T y_s, \quad
Ix+Jy_s \leq b \ [\lambda_s]
\end{align}

With $\lambda_s$ as the dual variable, the dual form of the SP is written as: 
\begin{align} \label{eq:dual}
    (\mathrm{dual \, SP}_s) \quad \max_{\lambda_s \geq 0}\lambda_{s}(b-Ix)\quad
s.t.\ \lambda_{s}J \leq \pi_s d^T
\end{align}

In our case, slack variables with extremely high costs ensure that the SPs are always feasible. Based on the strong duality theorem, the objective of the dual SP (Eq. \ref{eq:dual}) and of the SP (Eq. \ref{eq:SP}) coincide. Moreover, the feasible region of the SP is independent on the given $x$. Therefore, the maximization problem of dual SP in Eq. \ref{eq:dual} can be reformulated as a minimization problem: ${\mathrm{min}\ \{\tilde{z}_s(x)}| \lambda_{s} (b-Ix) \leq \tilde{z}_s(x)\}$. Substituting in the original MP (Eq. \ref{eq:MP}) yields the MP:

\begin{align} \label{eq:newMP}
    (\mathrm{MP}) \quad \min_{x \geq 0} c^T x  + \sum_{s\in S}\tilde{z}_s(x)\quad
s.t. \ Hx &\leq a, \quad
\lambda_{s} (b-Ix) \leq \tilde{z}_s(x)\  \forall s\in S 
\end{align}

The resulting constraint $\lambda_{s} (b-Ix) \leq \tilde{z}_s(x)$ is called an optimality cut. In BD, there are two variants of cut formulation: single-cut and multi-cut. The single-cut variant adds one cut aggregated across all SPs in each iteration to the MP. The multi-cut variant adds one cut per SP each iteration, as shown in Eq. \ref{eq:newMP}. Since the single-cut variant adds only a single constraint to the MP, it reduces the MP's solution time. The multi-cut variant adds more constraints, increasing the MP's solution time, but also provides more precise information for each SP, thereby reducing the number of iterations. Since the MP and its solution time are small in our case, the multi-cut formulation is more performative \citep{Jacobson2023}. 

\begin{figure}[!htp]
    \centering
        \includegraphics[scale=1.3]{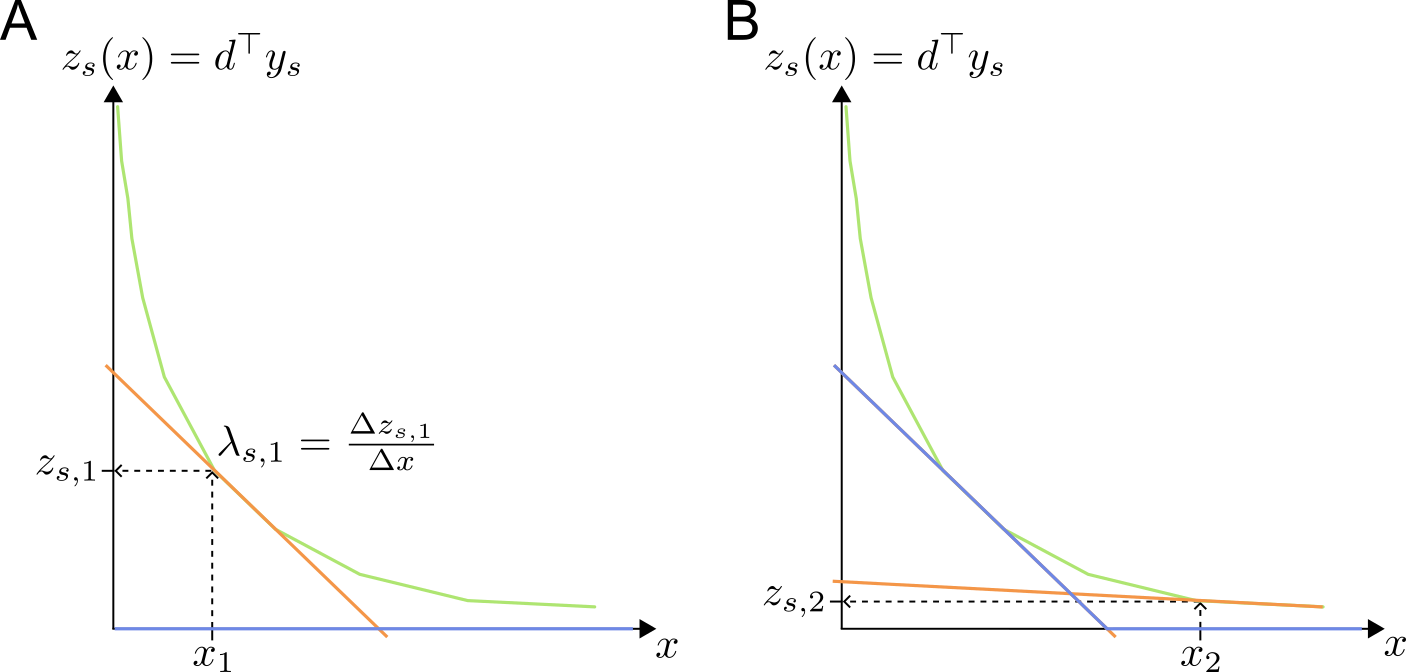}
    \caption{Cutting plane algorithm iteratively approximating the second-stage problems}
    \label{fig:cuttingplane1}
\end{figure}

\citet{firstenergybenders} provide a geometrical interpretation for the iterative cut addition of BD, visualized in Fig. \ref{fig:cuttingplane1}. The green line denotes the second-stage objective function in relation to a expansion variable $x$, and the blue line represents the current cutting plane estimator $\tilde{z}_s$, which evolves with each iteration to approximate the true second-stage objective function. The process unfolds as follows: 

\begin{enumerate}
    \item Initially, the cutting plane estimator is only constrained by non-negativity, indicated by the blue line in part (A) of Fig. \ref{fig:cuttingplane1}. 
    \item In the first iteration, solving the MP yields a candidate solution for the expansion variable $x_1$. Solving the SP and the dual SP with fixed $x_1$ provides the second-stage objective $z_{1,1}$ and the dual variable $\lambda_1$, where $\lambda_1$ can be interpreted as the derivative or sensitivity of $z(x)$ at the given $x_1$. Hence, the generated optimality cut can be rewritten as $\tilde{z}_s \geq z_s^{(i)} + \lambda_s^{(i)}(x - x^{(i)})$, represented by the orange line in part (A) of Fig. \ref{fig:cuttingplane1}.
    \item The newly added cut refines the cutting plane estimator, as represented by the blue line in part (B).
    \item In the subsequent iterations, the MP is solved with the updated cutting plane estimator, yielding new expansion variables $x_2$ and the estimated second-stage objective $\tilde{z}_{1,2}$. The SP is solved to obtain the actual second-stage objective $z_2$ at given $x$ and to generate the new cut, the orange line in part (B).
\end{enumerate}

Iteratively, the cutting plane estimator converges to the actual second-stage objective as a piece-wise linear function approximation. Since the second-stage objective function is convex, the cutting-plane estimator provides a valid lower bound on the second-stage objective. Thus, a lower bound $\zeta_{\mathrm{low}}$ for the original problem can be derived from the objective value of the MP. The upper bound $\zeta_{\mathrm{up}}$ is the best feasible result found $c^Tx+\sum_{s \in S}z_s$, where $z_s$ is the actual second-stage objective obtained from solving the SPs with the fixed $x$. The iterative process continues until the lower and upper bounds converge, meaning the relative optimality gap $1-\zeta_{\mathrm{low}}/ \zeta_{\mathrm{up}}$ is sufficiently small. Thus, the MP and the SPs need to be solved iteratively to generate and add more cuts, improving the accuracy of the cutting plane estimator, the quality of the MP solution, and the lower bound, and to find higher-quality solutions to improve the upper bound.

The general procedure for the standard BD is presented in Alg. \ref{alg:bd}. To summarize, BD iterates over 1) solving MP, 2) solving SP and adding cuts, 3) updating the lower and upper bound, until 4) the optimality gap is smaller than a predefined tolerance.

\begin{algorithm}
\caption{Basic Benders Decomposition}
\label{alg:bd}
\textbf{\textit{Step\ 0.1\ (Parameter initialization)}} \\
Set convergence tolerance $\varepsilon$, $\zeta_{\mathrm{up}} \gets \infty$,\  $\zeta_{\mathrm{low}} \gets 0$; \\
\textbf{\textit{Step\ 1\ (Solve MP)}} \\
Solve \textbf{MP} and get candidate solution $x^{(i)}$\; 
Set $\zeta_{\mathrm{low}} \gets {\mathrm{obj}(\textbf{MP})}$; \\
\textbf{\textit{Step\ 2\ (Solve SP)}} \\
$\forall s \in S$: Solve \textbf{$\mathrm{SP}_s$} with $x^{(i)}$ to get $z_s^{(i)}$ and \ $\lambda_s^{(i)}$; \\ 
Add cuts $\Omega_s^{(i)}(z_s^{(i)},\lambda_s^{(i)})$ to \textbf{MP}; \\
\textbf{\textit{Step\ 3\ (Update iteration)}} \\
\textbf{if} $c^T x + \sum_{s \in S} z_s < \zeta_{\mathrm{up}}$ \textbf{then} \\
 \quad $\zeta_{\mathrm{up}} \gets c^T x + \sum_{s \in S} z_s$ and set $x^{up} \gets x^{(i)}$; \\
\textbf{end if} \\
\textbf{\textit{Step\ 4\ (Convergence test)}} \\
\lIf {$1-\zeta_{\mathrm{low}}/ \zeta_{\mathrm{up}} < \varepsilon $}{stop}
\textbf{Return to \textit{Step 1}};
\end{algorithm}

\subsection{Regularized Benders decomposition} \label{regul}

Instability is a widely recognized drawback of basic BD \citep{Rahmaniani.2017}. The solutions to the MP in two consecutive iterations can be extreme points of the feasible region that are arbitrarily far apart. This issue is particularly pronounced in energy planning, where the feasible region is highly unrestricted and continuous \citep{Goke.2024}. Oscillations significantly slow convergence, as the cuts generated fail to guide the MP toward the optimal solution effectively. Bundle methods, also known as regularization or stabilization, can address this drawback, centering the iteration around a reference solution to prevent oscillation \citep{Frangioni.2002}.

Alg.\ref{alg:stab_bd} presents the general procedure for regularized BD. The algorithm requires an initial solution as a reference solution. In our case, we use the deterministic solution for the most probable scenario (Step 0.2). Solving the stable MP then determines the next candidate solution for the first-stage variables. Additionally, the original MP must be solved to obtain the true lower bound for the original problem (Step 1). In each iteration, if a new best solution is found, the algorithm updating the stability center to the current iterate, performing a \textit{serious step}; if not, a \textit{null step} is made ($q=0$), and the restrictiveness of the stabilization is potentially adjusted as necessary (Step 3).

Different regularization methods are applied to BD for energy planning to stabilize the MP \citep{Pecci.2025, zhang2024stabilised,Goke.2024}. The proximal-bundle method adds a penalty term to the objective function that penalizes moving away from the current stability center. The level-bundle method establishes a minimum-descent target to minimize deviation from the stability center, and the trust-region method restricts the next iteration to lie within a Euclidean hypersphere around the current stability center.

Although the three methods can in theory achieve the same convergence \citep{bonnans2006numerical}, the trust-region method proves to be the most robust in practice, as its performance is less sensitive to numerical configurations and requires less fine-tuning to achieve good results on specific problems \citep{Goke.2024}. Therefore, we apply this method throughout the paper, and our results are transferable to all other regularization methods.

\begin{algorithm}
\caption{Regularized Benders Decomposition}
\label{alg:stab_bd}
\textbf{\textit{Step\ 0.1\ (Parameter initialization)}} \\
Set convergence tolerance $\varepsilon$, $\zeta_{\mathrm{up}} \gets \infty$,\  $\zeta_{\mathrm{low}} \gets 0$; \\
\textbf{\textit{Step\ 0.2\ (Stability center initialization)}} \\
Solve deterministic problem for $s':= \{s \in S | \pi_s = \max(\pi_s)\}$; \\
Get $x^{(0)}$ and initialize stability center $x^{up} \gets x^{(0)}$;\\
\textbf{\textit{Step\ 1\ (Solve MP)}} \\
Solve regularized \textbf{MP}$'$ and get candidate solution $x^{(i)}$\; 
Solve original \textbf{MP} and set $\zeta_{\mathrm{low}} \gets  {\mathrm{obj}(\textbf{MP})}$; \\
\textbf{\textit{Step\ 2\ (Solve SP)}} \\
$\forall s \in S$: Solve \textbf{$\mathrm{SP}_s$} with $x^{(i)}$ to get $z_s^{(i)}$ and \ $\lambda_s^{(i)}$; \\ 
Add cuts $\Omega_s^{(i)}(z_s^{(i)},\lambda_s^{(i)})$ to regularized \textbf{MP}$'$ and original \textbf{MP}; \\
\textbf{\textit{Step\ 3\ (Update iteration)}} \\
\textbf{if} $c^T x + \sum_{s \in S} z_s < \zeta_{\mathrm{up}}$ \textbf{then} \\
 \quad $\zeta_{\mathrm{up}} \gets c^T x + \sum_{s \in S} z_s$ and set $x^{up} \gets x^{(i)}$; \\
 \quad Update regularizaiton of \textbf{MP}$'$; \\
\textbf{end if} \\
\textbf{\textit{Step\ 4\ (Convergence test)}} \\
\lIf {$1-\zeta_{\mathrm{low}}/ \zeta_{\mathrm{up}} < \varepsilon $}{stop}
\textbf{Return to \textit{Step 1}};
\end{algorithm}

\section{Prioritization of sub-problems}\label{ref2}

In the following, we introduce our surrogate-based approach for the prioritization of SPs. Since regularization is critical for the applications of BD in energy planning that we focus on, we develop a prioritization approach that is fully comptabile with regularization. Nevertheless, our method is equally applicable to BD without regularization and beyond energy planning.

Section \ref{3.1} presents the strategy for selecting the prioritized SP and establishing a benchmark to assess the potential improvement. In Section \ref{3.2}, we discuss options for surrogate models to approximate the SPs to support prioritization. Finally, we describe the complete workflow of surrogate-assisted sequential BD in Section \ref{3.3} and extend it to an asynchronous parallelization framework in Section \ref{3.4}. 

\subsection{Prioritization strategy} \label{3.1}
The idea of prioritization is to consider only the SPs that are most effective for convergence in each iteration. The challenge is to identify these SPs, as selecting random SPs risks increasing the total computational time since the increase in total iterations outweighs the decrease in time per iteration. Therefore, we prioritize the sub-problem with the largest current error of the cutting-plane estimator.

\begin{figure}[htpb]
\centering
\includegraphics[scale=1.3]{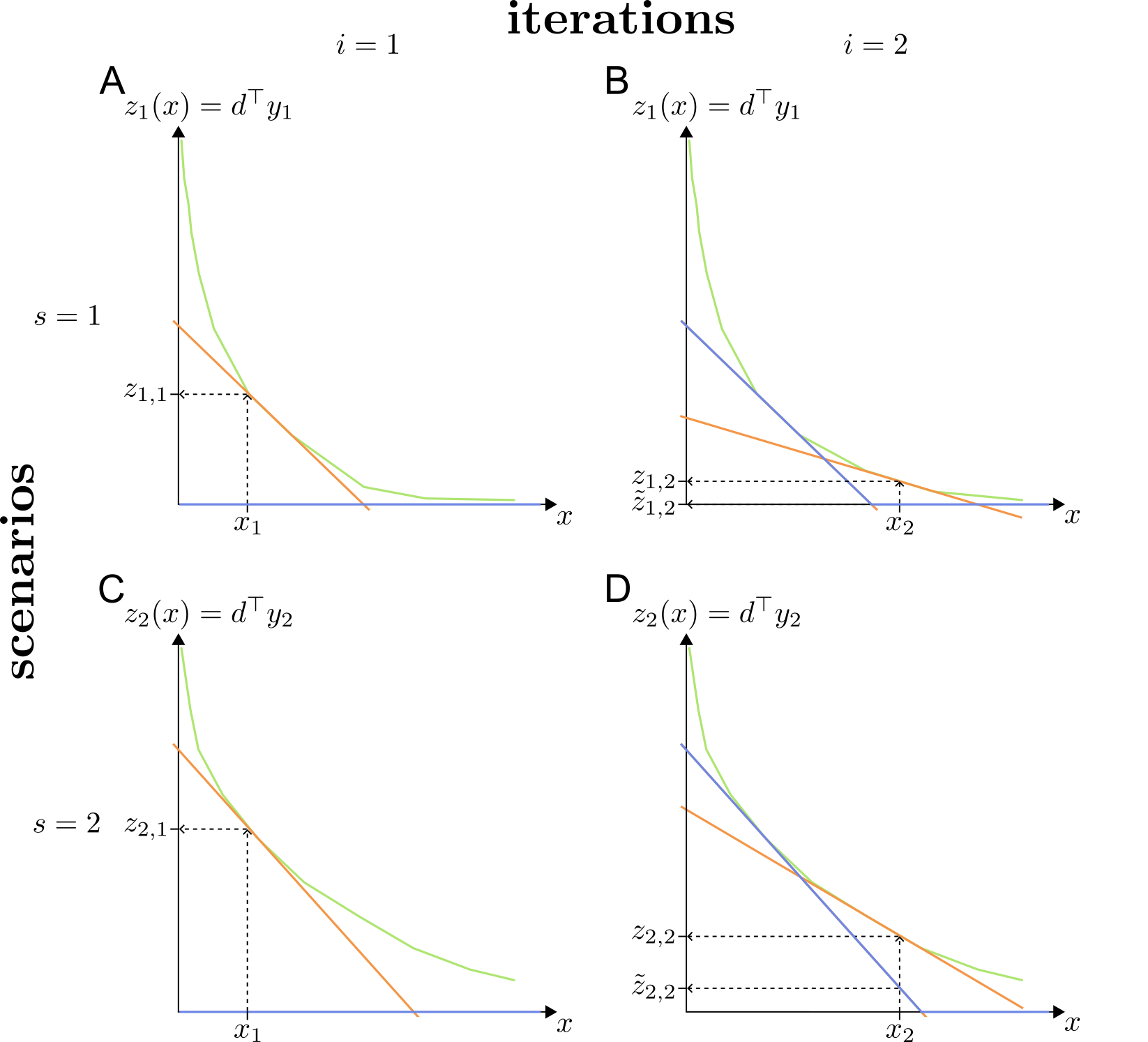}
\caption{Two SPs with different accuracy of cutting plane estimator in second iteration}
\label{fig:cuttingplane2}
\end{figure}

Fig. \ref{fig:cuttingplane2} visualizes shows the cutting-plane approximation of the SPs. In the figure, the difference $z -\tilde{z}$ corresponds to the current error of the cutting plane estimator for the respective SP. Comparing SPs 1 and 2 in the first and second rows, respectively, shows that the added cuts differ in their impact on the total error. In iteration $i=2$, the cutting plane estimator for SP 1 in part (B) of the figure already demonstrates good accuracy, making the cut added in that iteration less effective. In contrast, the cut for SP 2 in part (C) greatly improves the estimator's accuracy. As a result, computing the cut for the SP with the largest gap between $z_s$ and $\tilde{z_s}$ among all SPs is assumed to be more effective, while computing the other cuts is assumed to be expendable.

However, this idea has two caveats: First, the strategy is plausible, but it remains a heuristic criterion with no guarantee of strictly improving performance. Since solving fewer SPs alters the expansion variables obtained in subsequent iterations and, thus, the entire course of the algorithm. Therefore, identifying the SPs most effective for convergence with certainty is not even possible in hindsight. Generally, not solving all SPs introduces a trade-off: fewer SPs are solved per iteration, but more iterations are required to achieve convergence. Second, prioritizing which SPs to solve based on the difference between $z_s$ and $\tilde{z_s}$ assumes the actual second-stage objective $z_s$ are known \textit{before solving the SP}. However, in reality, exact information on $z_s$ requires solving the SPs in the first place.

\subsection{Surrogate-based approximation} \label{3.2}

Since the exact second-stage objective of each SP, $z_s$, is not available for the prioritization of SPs, we replace $z_s$ with the approximation $\hat{z}_s$ computed by a surrogate. Since the algorithms with prioritization described in the following sections \ref{3.3} and \ref{3.4} do not solve all SPs in each iteration, they depend on an approximation for the unsolved SP as a heuristic to decide if the current solution first-level $x^{(i)}$ should update the reference solution used for regularization \citep{de2014level,wolf2014applying}. Therefore, the approximation $\hat{z}_s$ will also be critical for updating the regularization. 

To compute the approximation, we use a surrogate that estimates the second-stage objective from the first-stage decision $x^{(i)}$. In our case, a challenge for the surrogate  is the high dimensionality of the first-level solution, corresponding to the number of expansion variables ($\approx100$), and the limited training data, corresponding to the previously solved instances of the SP during iteration. In addition, the surrogate must provide a more precise estimate than the cutting plane estimator $\tilde{z}_s$ to be sensible. But there are also several characteristics of the cutting plane estimator that are not required for the surrogate: the surrogate does not need to ensure feasibility, or provide lower bound and gradient information. 

In the following, we list conceivable surrogate methods, grouped into geometric interpolation and machine learning  methods:

\textbf{Geometric Interpolation} is used to estimate unknown values that fall within the range of a set of known data points. Interpolation constructs a function that closely matches the know data and provides reasonable estimates for points between the data. It is especially suitable for applications where the input space is too large to be explored adequately. Here, we introduce two methods:

\begin{itemize}

\item \textbf{Nearest neighborhood (NN)} assigns the value of the new point to that of the nearest known point. The interpolant is written as:
\[
    \hat{z}_{\mathrm{NN}}(x) = z(\underset{i}{\arg \min} \|x-x_i\|_2),
\]
where $x_i$ is the input for known points and $x$ is the input for the unknown point, $\|\ \|_2$ is the Euclidean distance, and z returns the known values. 

\item \textbf{Inverse Distance Weighting (IDW)} takes the weighted average of all known points, with weights inversely proportional to the distance between the new and known points.
\[
\hat{z}_{\mathrm{IDW}}(x) = 
\begin{cases}
\sum_{i=1}^n w_i(x)z_i, & \text{if } \|x-x_i\|_2 \neq 0 \\
z_i, & \text{if } \|x-x_i\|_2 = 0
\end{cases}
\]
where n is the number of known points, the weight $w_i$ is given by the inverse of distance:
\[
w_i(x) = \frac{1}{\|x-x_i\|_2^p}
\]

$p$ is the power parameter. Greater values of $p$ assign greater influence to values closest to the interpolated point.

\end{itemize}
\textbf{Machine Learning} is a popular option for surrogates, whose classic procedure is first to generate a large amount of data, then train a model to learn a mapping from inputs to outputs. Training data must be updated, and the surrogate retrained whenever variables, constraints, or parameters in the underlying optimization problem change. Accordingly, the training can only include data dynamically generated during previous iterations of the same computation. As a result, the number of input parameters far exceeds the number of training data points, and thus, we only considered machine learning methods that are more suitable for high-dimensional inputs:
\begin{itemize}
    \item \textbf{Lasso Regression} is a type of linear regression that includes a penalty term proportional to the absolute value of the coefficients. This penalty helps enforce sparsity in the model, incentivising to select a subset of relevant features.\\
  \item \textbf{Support Vector Regression (SVR)} utilizes kernel functions to approximate the target values within an acceptable error margin. This makes SVR less sensitive to extreme values at the earlier iterations. One choice for the kernel function is the radial basis function, which is also often used in interpolation. The approximation is written as a linear combination of functions depending on the distance between two points:
    \[
    \hat{z}_{\mathrm{RBF}}(x) = \sum_{i=0}^{N-1} w_i\phi(\|x-x_i\|_2)
    \]
    The interpolating function passes through all known points, yielding $w_i$ that minimize the error. 
    \item \textbf{Multi-Layer Perceptron (MLP)} is the foundation of Artificial Neural Network, containing one or more hidden layers. Neural Networks require large training data sets. However, since our training dataset is relatively small, we restrict our MLP to a single hidden layer to avoid overfitting.
\end{itemize}

\subsection{Prioritization in sequential Benders decomposition} \label{3.3}

\begin{figure}
    \centering
    \includegraphics[scale=1.0]{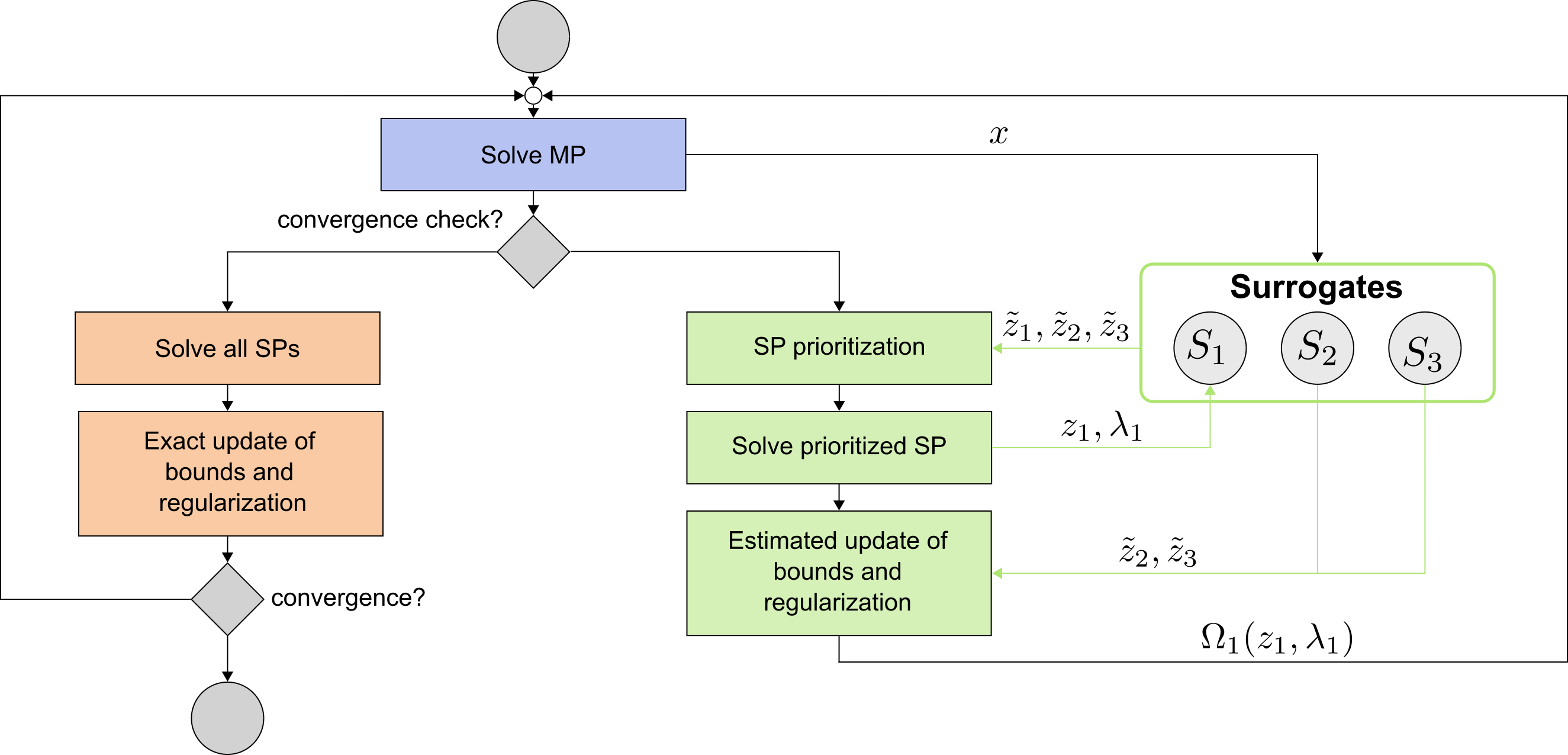}
    \caption{Flowchart for sequential implementation of SP-prioritization}
    \label{fig:seq}
\end{figure}

Fig. \ref{fig:seq} and Alg. \ref{alg:serial_bd} describe how to implement SP-prioritization into sequential regularized BD. As motivated in previous sections, instead of solving all SPs in each iteration, we first deploy a surrogate to compute $\hat{z}_s$ as an estimate of the second-stage objective value. Then we select the SPs with the largest $\hat{z}_s-\tilde{z}_s$ and solve only those SPs, since we expect their cuts to be most effective for convergence. Later, these exact SP solutions update the surrogate models. 

To update the regularization and bounds, we again use the surrogate to estimate the second-stage objective value of the unsolved SPs. Since the estimates are neither exact nor provide a valid bound, relying solely on the surrogate cannot guarantee convergence. The underestimation of the second-stage objective by the surrogates may result in one SP being left unsolved, even though convergence requires solving it. For the same reason, the SP-prioritization can result in the stability center being stuck at a suboptimal solution.

\begin{algorithm}[H]
\caption{Sequential and regularized BD with SP-prioritization}
\label{alg:serial_bd}
\textbf{\textit{Step\ 0.1\ (Parameter initialization)}} \\
Set convergence tolerance $\varepsilon$ and distance threshold $\eta$, $\zeta_{\mathrm{up}} \gets \infty$,\  $\zeta_{\mathrm{low}} \gets 0$; \\
\textbf{\textit{Step\ 0.2\ (Stability center initialization)}} \\
Solve deterministic problem for $s':= \{s \in S | \pi_s = \mathrm{max}(\pi_s)\}$; \\
Get $x^{(0)}$ and initialize stability center $x^{up} \gets x^{(0)}$;\\
\textbf{\textit{Step\ 1\ (Solve MP)}} \\
Solve regularized \textbf{MP}$'$ and get candidate solution $x^{(i)}$\; 
Solve original \textbf{MP} and set $\zeta_{\mathrm{low}} \gets {\mathrm{obj}(\textbf{MP})}$; \\
\textbf{\textit{Step\ 1\ (Solve SPs)}} \\
\textbf{if} $\|x_i-x_{i-1}\|_2 \leq \eta$  \textbf{or}  $1-\zeta_{\mathrm{low}}/\zeta_{\mathrm{up}} \leq \varepsilon$  \textbf{then} \\
\quad $\forall s \in S$: Solve \textbf{$\mathrm{SP}_s$} with $x^{(i)}$ to get $z_s^{(i)}$ and \ $\lambda_s^{(i)}$; \\ 
\quad set $c \gets true$; \\
\textbf{else}  \\
\quad $\forall s \in S$: compute surrogate $\hat{z}_s$; $S_{p} = \underset{s}{\arg \max} (\hat{z}_s-\tilde{z}_s)$ \\
\quad $\forall s \in S_p$: Solve \textbf{$\mathrm{SP}_s$} with $x^{(i)}$ to get $z_s^{(i)}$ and \ $\lambda_s^{(i)}$; \\ 
\quad Train surrogate model for $\mathrm{SP}_s$ with $z_s^{(i)}$ for $s \in S_p$; \\
\quad set $c \gets \mathrm{false}$; \\
\textbf{end if} \\
Add cuts $\Omega_s^{(i)}(z_s^{(i)},\lambda_s^{(i)})$ to regularized \textbf{MP}$'$ and original \textbf{MP}; \\ 
\textbf{\textit{Step\ 3\ (Update Iteration)}} \\
\textbf{if} $\|x_i-x_{i-1}\|_2 \leq \eta$  \textbf{or}  $1-\zeta_{\mathrm{low}}/\zeta_{\mathrm{up}} \leq \varepsilon$  \textbf{then} \\
    \quad $x^{up} \gets x^{(i)}$;\\
\textbf{end if} \\
\textbf{if} $c^T x + \sum_{s \in S} z_s < \zeta_{\mathrm{up}}$ \textbf{then} \\
 \quad $\zeta_{\mathrm{up}} \gets c^T x + \sum_{s \in S} z_s$ and set $x^{up} \gets x^{(i)}$; \\
 \quad Update regularizaiton of \textbf{MP}$'$; \\
\textbf{end if} \\
\textbf{\textit{Step\ 4\ (Convergence Test)}} \\
\lIf {$1-\zeta_{\mathrm{low}}/ \zeta_{\mathrm{up}} < \varepsilon $ \textbf{and} $c$}{stop}
\textbf{Return to \textit{Step 1}};
\end{algorithm}

To address the problem of halted convergence, we introduce a regular convergence check based on two conditions:
\begin{enumerate}
    \item \textbf{Convergence of the approximated upper bound}: This condition is met, if the surrogate-based upper bound and the lower bound are below a convergence tolerance. In the ideal case, where the surrogate provides an accurate estimate, the algorithm converged.
    \item \textbf{Proximity of complicating variables}: This condition is met, if the complicating variables in the current iterations $x^{(i)}$ are sufficiently close to values from the previous iteration $x^{(i-1)}$, suggesting convergence halted. Invoking a convergence check at this point prevents repeatedly solving ineffective SPs while overlooking others that were not solved for an extended period. To quantify the distance between two consecutive expansion variables, we use the Euclidean distance and define a threshold share $\eta$ in relation to the absolute distance of the current complicating variables: $\eta = h \cdot \|x_i\|_2$, where $h$ is a constant threshold distance parameter.
\end{enumerate}

If the conditions for a convergence check hold, the current iteration solves all SPs. Furthermore, a convergence check forces an update of the stability center for regularization.

\subsection{Prioritization in parallelized Benders decomposition}  \label{3.4}

\begin{figure}[htb]
\centering
\includegraphics[scale=0.8]{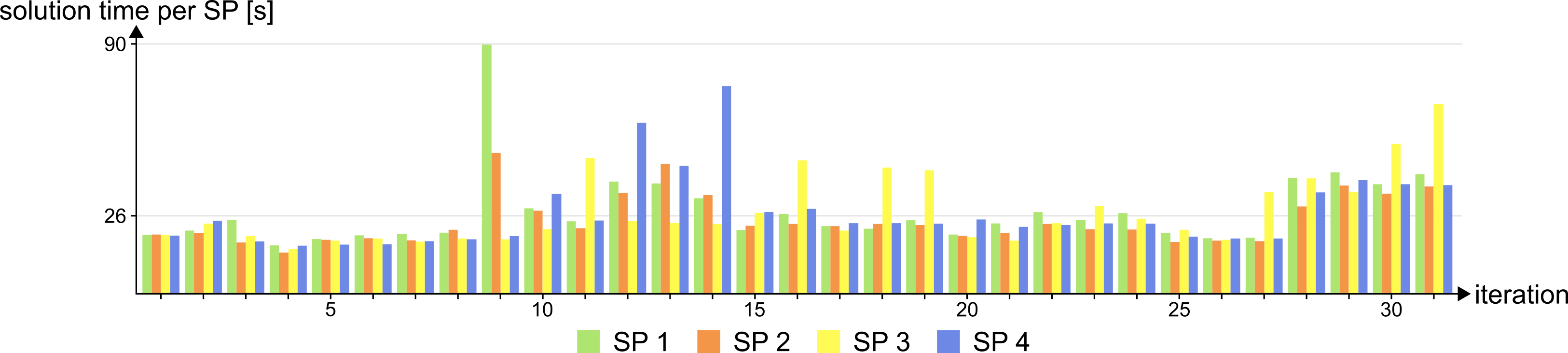}
\caption{Solution time for SPs in case with 4 scenarios}
\label{fig:timeSub}
\end{figure}

Instead of solving the SPs sequentially, the parallelized BD solves them in parallel, enabling distributed computing across multiple workers running on different nodes. However, in the standard parallelized BD shown in part (A) of Fig. \ref{fig:timeSub}, the solution time of the SPs can vary significantly due to random fluctuations in solution time and differences in the technical characteristics of the computing nodes. Since re-solving the MP waits for all SPs to finish, the parallelized BD suffers from high idle time and cannot efficiently utilize available computing resources. To reduce idle time, asynchronous parallelization does not wait for all SPs to finish until re-solving the MP. Accordingly, the unsolved SPs will not add cuts for the next solve of the MP, potentially reducing the quality of the next candidate solution and, thus, increasing the number of iterations.

\begin{figure}[htb]
\centering
\includegraphics[scale=1.0]{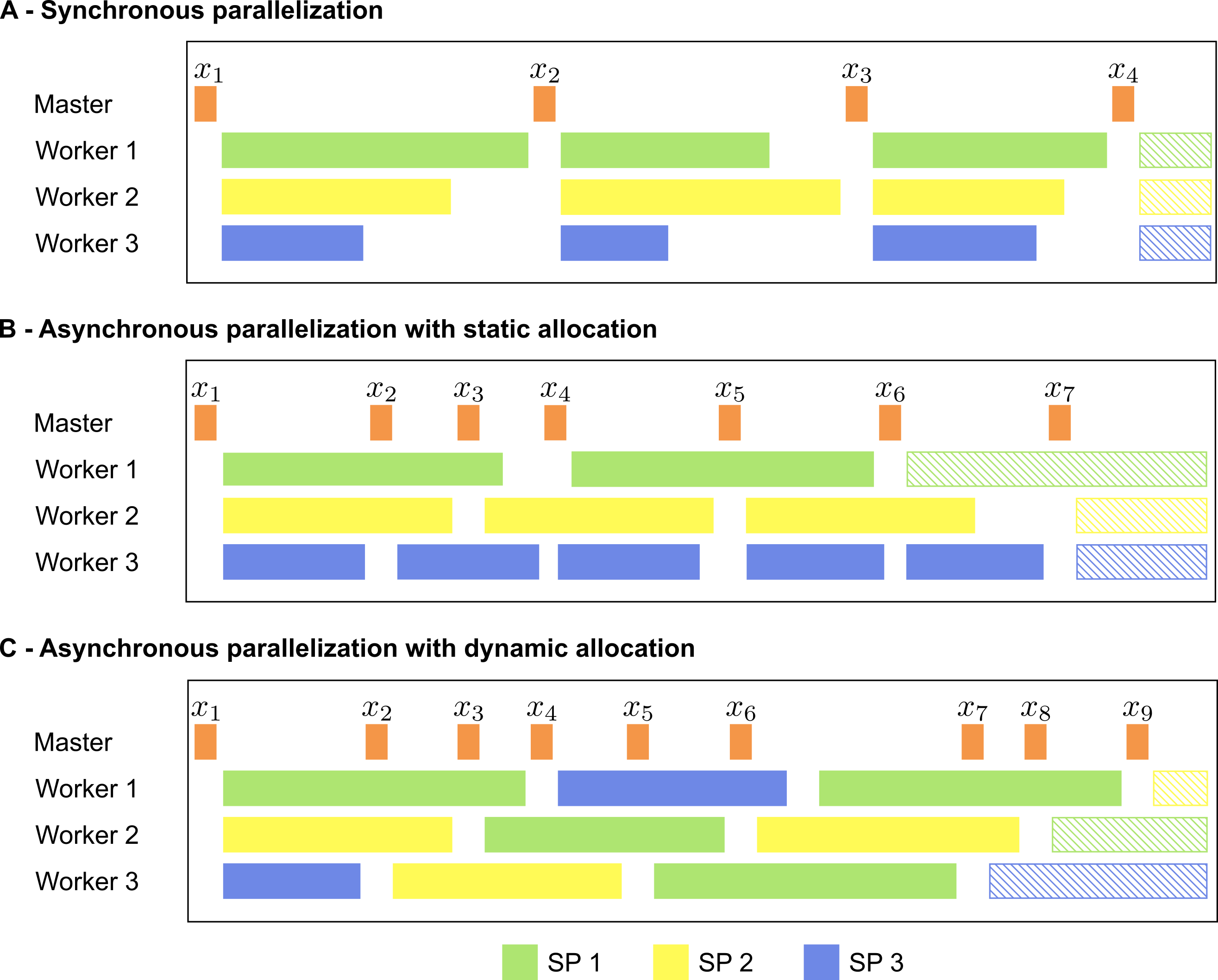}
\caption{Concept of synchronous and asynchronous parallelization}
\label{fig:conceptP}
\end{figure}

In this paper, we introduce two surrogate-based asynchronous variants of parallelized BD: 
\begin{enumerate}
\item \textbf{Static allocation} assigns exactly one worker to each SP. Accordingly, every worker must initialize only one SP and can run only one SP, too. Instead of waiting for all SPs to finish before re-solving the MP, this variant waits only for the prioritized SP. The prioritized SP is again determined by estimating second-stage costs using a surrogate and computing $\hat{z}_s-\tilde{z}_s$. When a non-prioritized SP finishes, it is either restarted with the latest candidate solution from the MP or waits for the MP to compute a new candidate solution.
\item \textbf{Dynamic allocation} assigns all workers to all SPs. Thus, every worker must initialize all SPs, which increases pre-processing time, but enables more flexibility when running SPs. This variant resolves the MP and updates the prioritization whenever any SP finishes. The freed worker directly starts the SP with the highest priority for the latest candidate solution. Prioritization again uses the surrogate-based approach.
\end{enumerate}
Fig. \ref{fig:conceptP} illustrates the difference between standard synchronous parallelization and the two introduced variants of surrogate-based asynchronous parallelization. The figure schematically shows that dynamic allocation is expected to achieve the least idle time, but at the expense of increased pre-processing time to initialize all SPs on each worker.

The two variants of asynchronous parallelization apply rules similar to those of the sequential BD to guarantee convergence when using prioritization. If the distance between the current candidate solution and the last candidate solution that a SP was solved for is above a predefined threshold, the SP is assigned a high priority; vice versa, if the distance between the current candidate solution and the last candidate solution that a SP was solved for is below a certain threshold, the SP is assigned a low priority. Both thresholds are defined relative to the absolute distance of the current candidate solution, using the threshold parameters $H_{max}$ and $H_{min}$, respectively.

Unlike synchronous parallelization, asynchronous parallelization does not solve every SP for every candidate solution and, thus, cannot provide the second-stage objective that provides an exact upper bound for the convergence check and the update of the reference point for the regularization. Therefore, the asynchronous parallelization uses the surrogates to estimate the objective of unsolved SPs, analogously to the sequential algorithm in the previous section. If the estimated objective indicates that the optimality gap is below the convergence tolerance, all SPs are solved for the corresponding complicating variables to confirm the convergence. If the estimated objective indicates that the candidate solution improves the current best, the reference point for the regularization is updated accordingly.

A key difference between synchronous and asynchronous parallelization is the communication between the master solving the MP and the workers solving the SPs. In synchronous parallelization, the MP simply waits for all SPs to finish. In asynchronous parallelization, the algorithm coordinates resolving the MP and the SPs with evaluating their results to prepare for the next computations.

\begin{figure}[htb]
\centering
\includegraphics[scale=1.0]{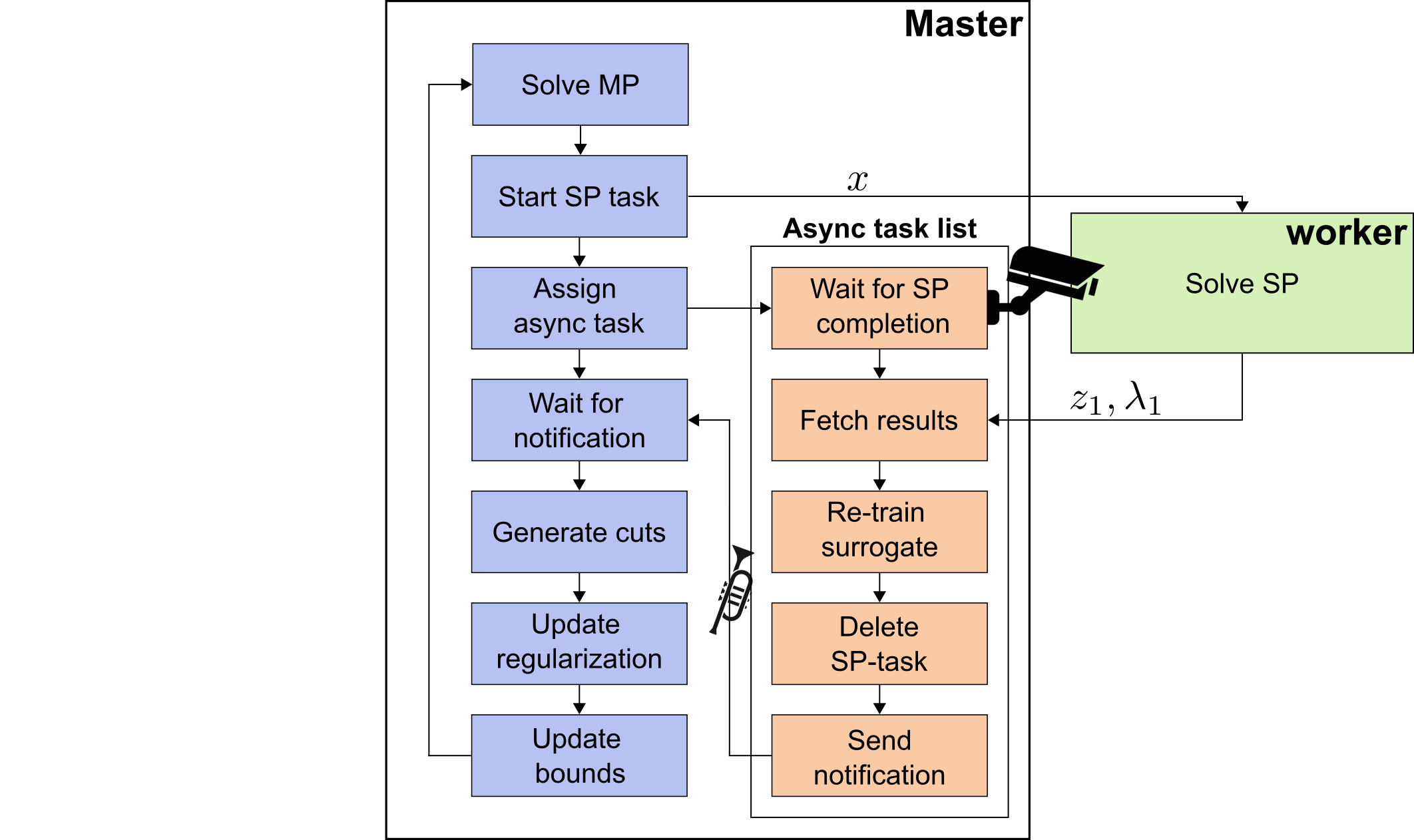}
\caption{Communication mechanism and workflow of asynchronous parallelization}
\label{fig:communication}
\end{figure}

To coordinate these tasks, we apply the communication workflow outlined in Fig. \ref{fig:communication}. After solving the MP, the algorithm starts a new SP and then updates a list of asynchronous tasks that run concurrently to the other operations on the worker for the MP. The tasks include monitoring workers' statuses, retrieving results for solved SPs, and retraining the surrogate based on the results. Afterward, the task list informs the main process of the SPs' completion. When the master task receives the info of a finished SP, it updates the cuts, stability center, and bounds accordingly, and resolves the MP. 

\section{Case study} \label{case}

\begin{figure}[htb]
\centering
\includegraphics[width=0.8\textwidth]{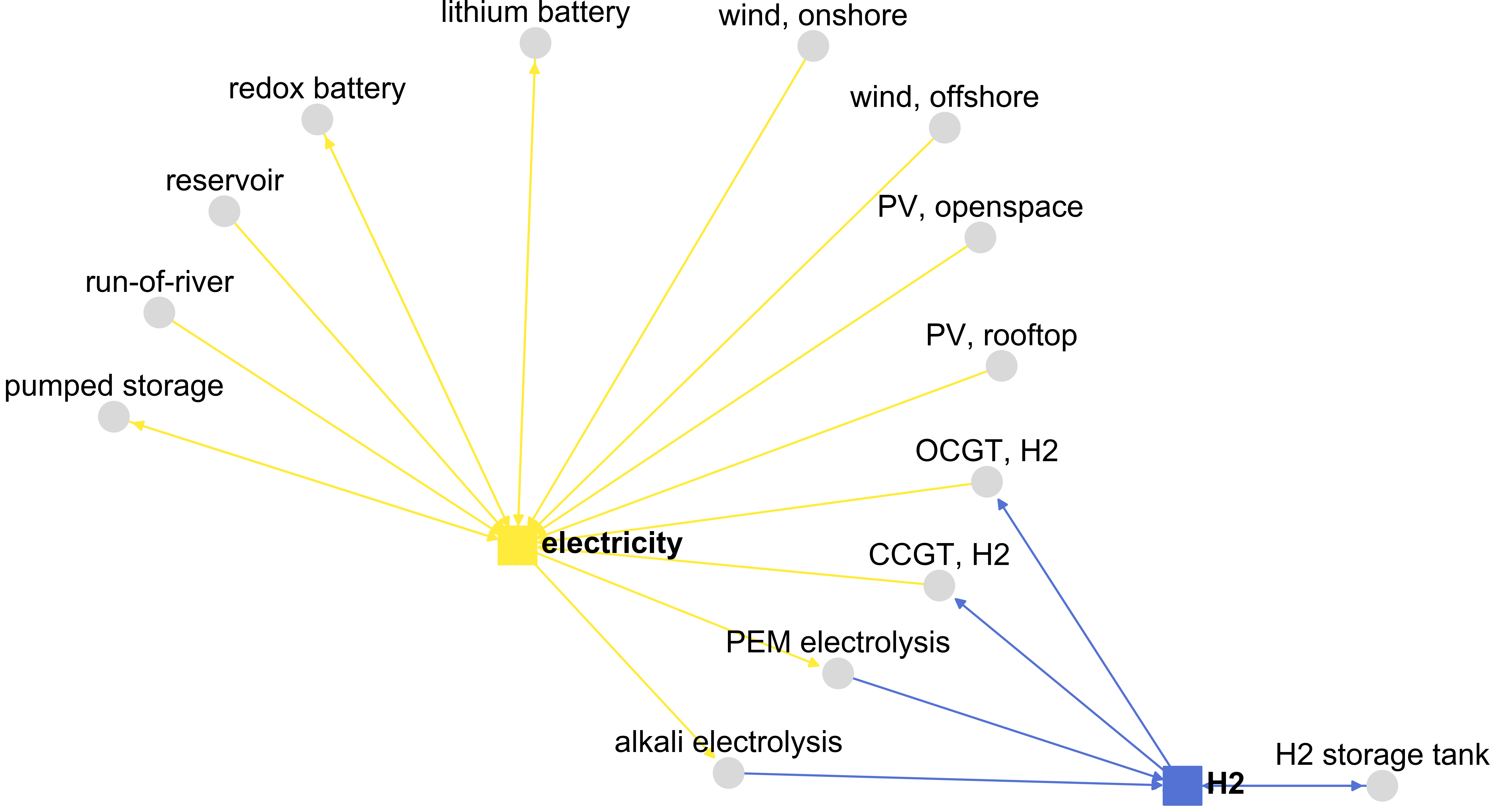}
\caption{Overview of energy carriers and technologies in the capacity expansion model}
\label{fig:conversion}
\end{figure}

This section introduces the case study used to benchmark the algorithms introduced previously. The applied capacity expansion model uses a greenfield approach to plan a fully decarbonized power system. The considered energy carriers, shown as colored squares in Fig. \ref{fig:conversion}, include electricity and hydrogen. In the graph, grey nodes represent technologies, the directed edges between nodes represent possible paths for energy conversion and storage, and squares describe possible paths for energy conversion and storage. The exogenous demand that the model must meet is limited to electricity. Hydrogen is included to enable long-term storage of electricity via electrolysis, storage tanks, and hydrogen turbines, in addition to short-term storage of electricity with batteries.

Spatially, the model covers four nodes, each corresponding to one country: France, Belgium, the Netherlands, and Germany. Temporally, supply and demand for electricity are modeled at hourly resolution over an entire year, yielding 8,760 time steps. Hydrogen uses a daily resolution. For details on implementing different temporal resolutions, see \citet{Goeke2020b}. Different scenarios for the parameters describing electricity demand, wind, and solar energy supply, corresponding to historical climate years, introduce second-stage uncertainty.

The expansion variables in the model cover both the transmission and generation capacity. Wind and solar capacity are subject to an upper limit that restricts expansion. Thermal power plants include combined cycle and open cycle gas turbine, OCGT and CCGT, respectively, fuelled by hydrogen. For hydro power, viz. run-of-river, pumped storage, and reservoirs, the total capacity is fixed at current values, since the literature typically assumes that the potential for hydro power in Europe is already exhausted \citep{Horsch2018}. The total number of expansion variables is approximately 100.

\section{Results} \label{bench}

The first section of the results verifies the idea underlying our prioritization strategy that a cut's impact on overall convergence correlates with its improvement of the cutting-plane estimator for the current candidate solution. Next, in Section \ref{acSur}, we compare different surrogates to estimate this improvement without actually solving the SP. Afterward, we deploy the most promising surrogates to benchmark the sequential and asynchronous algorithm with SP-prioritization in Section \ref{seqBd2} and \ref{seqBd2}, respectively.

For parallelized BD, the MP and SPs for each scenario run on independent computing cluster nodes, each with four cores and 16 GB of memory. Since the technical specifications of the cores can differ across runs and nodes, we reran the unexpected performance outliers to ensure the robustness of the results. For solving problems, each node deploys the Barrier implementation of Gurobi 10.1 without crossover and the \texttt{NumericFocus} parameter set to zero. For BD, we delete unused cuts after 20 iterations and choose a convergence tolerance $\varepsilon$ of 0.1\%. All deployed variations of BD are implemented in the version of the AnyMOD.jl modeling framework linked in the Supplementary material. Generally, total run-time depends on the number of iterations and the average time to solve the SPs since the MP is small.

\subsection{Impact of SP-prioritization} \label{spPrio}

As a first step, we verify that prioritizing the SP with the greatest improvement of the cutting-plane estimator for the current candidate solution improves performance. In this benchmark, we compare three cases: \textit{Standard} is the standard regularized BD, which solves all SPs in each iteration. The other two cases solve only one SP in each iteration. \textit{Prioritization}  prioritizes the SP with the largest $z_s - \tilde{z}_s$ and \textit{Random} chooses a random SP. For the \textit{Prioritization} case, we assume we have a perfect surrogate and know the exact second-stage objective $z_s$ for each SP before solving it. In practice, we simulate having this information by still solving all SPs, but using only the prioritized SP to generate cuts and update bounds.

\begin{figure}[htpb]
        \centering
        \includegraphics[scale=0.8]{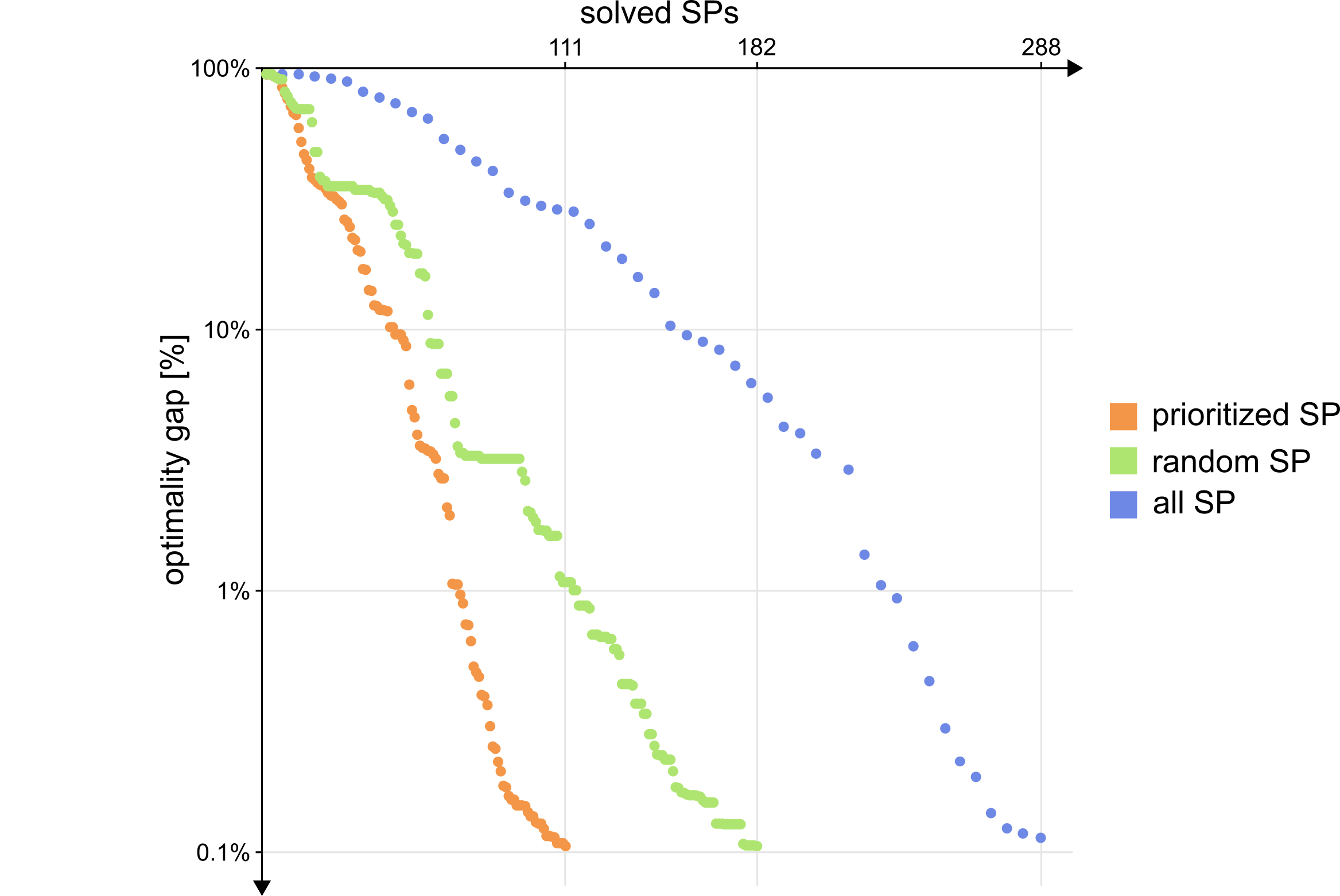}
    \caption{Benchmarks for SP prioritization (6 scenarios)}
    \label{fig:SPprior}
\end{figure}

Fig. \ref{fig:SPprior} shows the performance of three cases with six second-stage scenarios. As a performance metric, we use the number of SPs solved, since the MP solution time is negligible in our case, consistently accounting for less than 1\% of the total solution time. Furthermore, the number of SPs solved reflects the benefit of solving fewer SPs per iteration, as well as the adverse effect of needing more iterations for the algorithm to converge.

The results indicate that our strategy for SP-prioritization is sensible. With standard BD, the algorithm solves 288 SPs in 48 iterations, corresponding to the number of blue points in Fig. \ref{fig:SPprior}; when solving only the prioritized SP, it converges after 111 iterations, solving 111 SPs. Solving a single random SP instead of the prioritized SP increases the number of solved SPs to 182, confirming that the benefits come from prioritization rather than from solving a single SP in each iteration.

\subsection{Accuracy of surrogate models} \label{acSur}

Next, we evaluate the surrogate methods for estimating the second-stage objective. For testing, we use the values for the complicating variables and the corresponding second-stage objective values for six SPs and 37 iterations. Since the standard BD converges in 38 iterations for this problem, the tested sample reflects the typical range of values a surrogated would be applied on.

We start with the first 10 observations for training, and successively increase the training set by one observation, emulating training during the iterations of the algorithm, up to a maximum of 35. The remaining observations are used for post-training out-of-sample testing.

The surrogate methods introduced in Section \ref{3.2} use the following configurations:
\begin{itemize}
    \item \textbf{Lasso:} The Lasso regression is implemented with a regularization parameter set to 0.5. This parameter controls the amount of shrinkage applied to the coefficients, thereby mitigating overfitting while promoting feature selection.
    \item \textbf{SVR:} The SVR model utilizes a RBF kernel, configured with hyperparameters $c$ set to 1.0 and epsilon to 0.1. The hyperparameter $C$ controls the trade-off between a smooth decision boundary and correctly classifying training points, while epsilon determines the width of the epsilon-insensitive tube around the regression function.
    \item \textbf{MLP:} The MLP consists of an input layer with the size of input parameters and a hidden layer with 5 neurons, both employing ReLU activation functions. The model is trained using the Adam optimizer for 10 epochs with a batch size of 16.
    \item \textbf{IDW:} The IDW uses a power parameter of 5.
    \item \textbf{NN:} No parameter requires.
\end{itemize}

\begin{figure}[htbp]
\centering\includegraphics[scale=0.8]{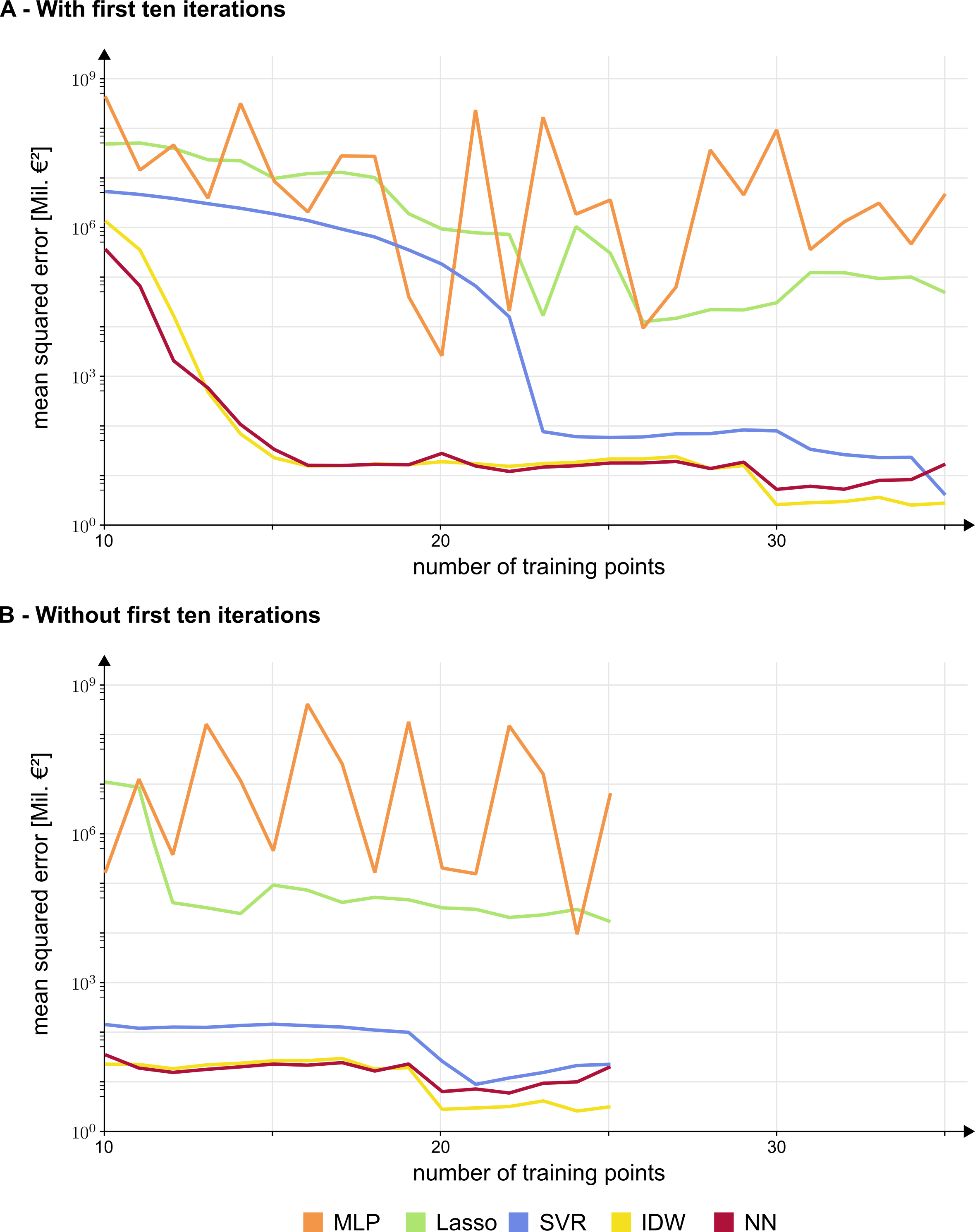}
\caption{Accuracy of surrogates}
\label{fig:sur1}
\end{figure}

Fig. \ref{fig:sur1} shows the mean squared error of the surrogates to measure their accuracy. As mentioned in Section \ref{3.2}, the objective values in the early iterations of BD are exceptionally high, significantly distorting the performance of machine learning methods. Therefore, part (B) excludes these first ten iterations and instead trains on iterations 10 to 20.

Both variations show that the geometric interpolation methods, namely IDW and NN, significantly outperform the machine learning method and achieve a strong correlation between the predicted and actual values. Consequently, omit the machine learning methods in the following sections.

\subsection{Performance with sequential BD} \label{seqBd}

\begin{figure}[htb]
\centering
\includegraphics[scale=1.0]{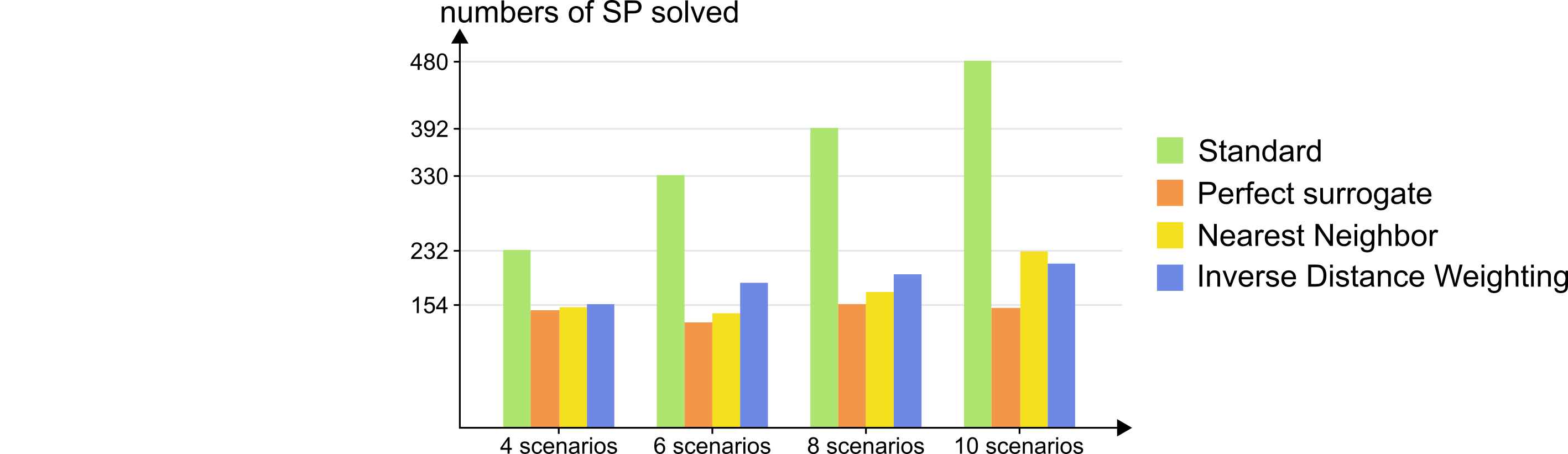}
\caption{Performance of sequential implementation}
\label{fig:seq2}
\end{figure}

To benchmark the performance of the interpolation-based surrogates, Fig. \ref{fig:seq2} 
compares the number of SPs solved to reach convergence. The threshold distance parameter $h$ for the convergence check is set to 10 and for regularization, the benchmark uses the trust-region method.

The benchmark distinguishes four cases: First, \textit{Standard} is the algorithm without prioritization, solving all SPs in each iteration. The other cases solve only one prioritized SP per iteration. \textit{Perfect surrogate} is a hypothetical case, assuming a surrogate perfectly predicts the second-stage objectives. In practice, we simulate the perfect surrogate by still solving all SPs, but using only the prioritized SP to generate cuts and update bounds, as in Section  \ref{spPrio}. Finally, the cases \textit{Nearest Neighbor} and \textit{Inverse Distance Weighting} use the respective interpolation methods to prioritize SPs.

The results show that SP-prioritization significantly reduces computational effort. Since prioritization solves only one SP instead of all SPs, and each SP corresponds to one scenario, the number of avoided computations increases with the number of considered scenarios. Accordingly, the reduction also increases with the scenarios. With 4 scenarios, SP-prioritization reduces the number of solved SPs by 33\%; with 10 scenarios, the reduction increases to 55\%. All cases successfully converge to the predefined tolerance of 0.1\%.

On average, Nearest Neighbor performs slightly better than Inverse Distance Weighting. For small scenario numbers, both surrogates are almost as good as the perfect surrogate. Still, as the number of scenarios increases and prioritization must select from a larger number of SPs, the gap widens.

\subsection{Performance with parallelized BD} \label{seqBd2}

The benchmark of parallelized BD compares the three cases already introduced in Section \ref{3.4} and Fig. \ref{fig:conceptP}: \textit{Standard} refers to the synchronous parallelization approach, \textit{Static} is the asynchronous parallelization with static allocation, assigning exactly one worker to each SP, and \textit{Dynamic} assigns all workers to all SPs. The computations run on a distributed-memory system, with each worker corresponding to a node with 4 cores and 16 GB of memory. The threshold distance parameters $H_{\mathrm{min}}$ and $H_{\mathrm{max}}$ for the \textit{Dynamic} case are set to 0.0001 and 5, respectively. The \textit{Static} and \textit{Dynamic} cases use the nearest neighbor as a surrogate.

Benchmarking the parallelized setups is complex due to the lack of a conclusive metric. In the parallelized setup, the number of solved SPs does not account for the reduction in idle time. Total solution time remains highly variable because the allocated hardware can differ between runs, and the solvers used for the SPs and MPs are non-deterministic. Therefore, we use both metrics to provide an exhaustive overview and report median values for the total solution times.

\begin{figure}[htb]
\centering
\includegraphics[scale=1.0]{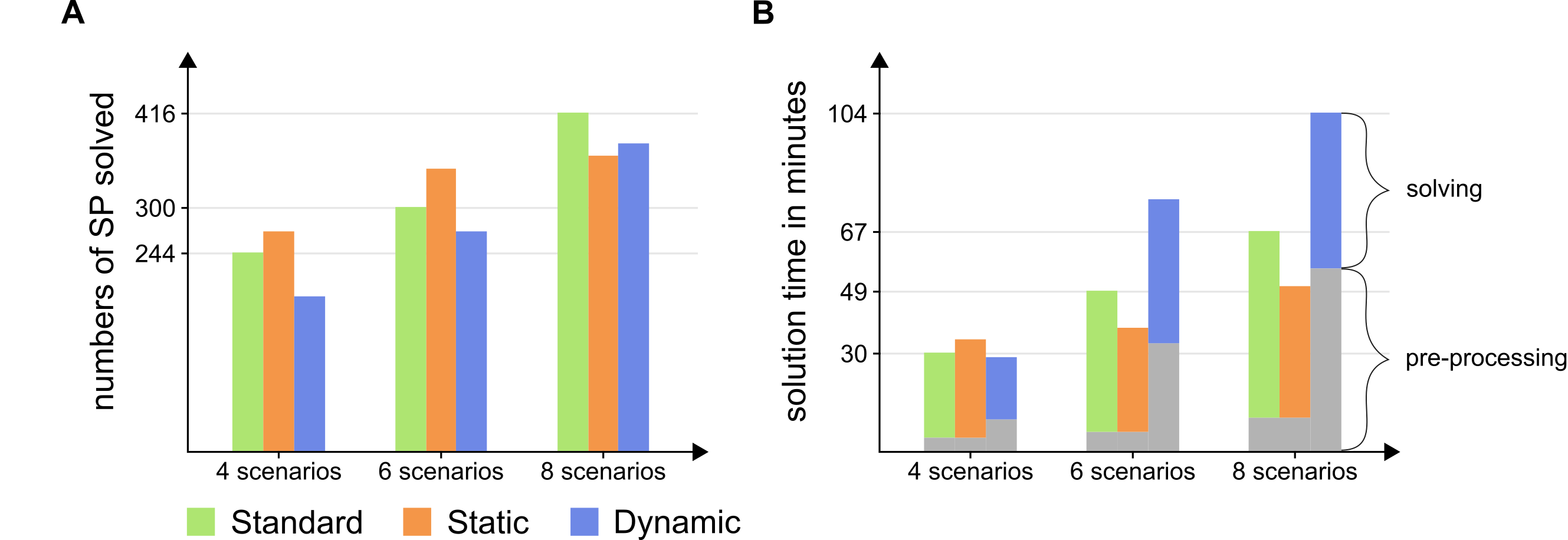}
\caption{Number of SPs solved and median solution time for parallelization methods}
\label{fig:asySP}
\end{figure}

Part A of Fig. \ref{fig:asySP} compares the number of solved SPs across the three parallelized cases. Asynchronous parallelization with static SP-prioritization increases the number of solved SPs in two cases and decreases the number of solved SPs in one case. In contrast, dynamic allocation consistently reduces the number of iterations. This difference reflects that the dynamic variant can more flexibly prioritize critical SPs, since each worker can run any SP. As a result, dynamic allocation tends to yield more valuable cuts to the MP, leading to fewer iterations.

To complement the results on solved SPs, part B of Fig. \ref{fig:asySP}  compares the total solution time across the three parallelized cases. Although static allocation is disadvantageous in terms of the number of solved SPs, thanks to reduced idle time in asynchronous parallelization, static allocation can achieve significant reductions in solution time. With 4 scenarios, there is still a slight increase in solution time by 11\%, but with 6 scenarios, static allocation reduces solution time by 29\%, and with 8 scenarios, by 36\%. 

In addition, the analysis of solution times demonstrates the major disadvantage of dynamic allocation. While the strategy appeared promising in terms of the number of solved SPs, these benefits are offset by a substantial increase in pre-processing time required to initialize each SP on each worker.

\section{Conclusion}

This work introduces surrogate-based prioritization as a refinement for BD. This method leverages surrogates to estimate the sub-problems' objectives, assesses the current error of the cutting-plane estimator, and then prioritizes the sub-problem with the largest error. We develop one sequential and two asynchronous Benders algorithms with surrogate-based prioritization that can also apply regularization.

We benchmark the introduced methods on a large-scale capacity expansion problem in energy planning. The results show that reductions in solution time correlate with surrogate accuracy and that surrogate-based prioritization outperforms random prioritization. In our test case with a small training set, interpolation methods outperform machine learning as a surrogate in terms of prediction error. In the sequential algorithm, surrogate-based prioritization consistently and significantly reduces computation, and the speed-up increases with the number of scenarios. With asynchronous parallelization, the results are less clear, and performance suffers due to the high preprocessing time required to initialize the SPs on all workers.

The proposed concept of SP-prioritization is generally transferable to other problems solved with BD. The application is sensible whenever solving the SPs consumes a significant share of the solution time; even if sample-average approximation for SP selection is applicable, both methods can be combined. Transferring the approach to other problems suggests adjusting the surrogate method used. Depending on the structure, size, and number of SPs, machine-learning methods paired with pre-training or cross-SP training may outperform geometric interpolation methods.

In addition, SP-prioritization can contribute to the development of sophisticated asynchronous algorithms. Our work shows that the surrogate-based approach can consistently identify the SPs that provide the greatest benefit to convergence, but that leveraging this information in a parallelized algorithm is complex. Therefore, future work should further investigate trade-offs in allocating SPs to workers in distributed memory systems. Furthermore, SP-prioritization could be paired with machine-learning-based estimates of solution time to support a rigorous cost-benefit-based parallelization scheme.

\section{Acknowledgments}

We would like to thank André Bardow and Stefano Moret for their valuable feedback on this work.

\section{CRediT authorship contribution statement}

\textbf{Wanhong Yu}: Conceptualization, Data curation, Investigation, Methodology, Software, Validation, Visualization, Writing – original draft, Writing – review \& editing.
\textbf{Boyung Jürgens}: Conceptualization, Validation, Writing – review \& editing.
\textbf{Leonard Göke}: Conceptualization, Validation, Supervision, Writing – review \& editing.

\printcredits


\bibliographystyle{cas-model2-names}
\bibliography{cas-refs}


\end{document}